\documentclass{arxiv-bioinfo}
\pdfoutput=1

\usepackage{mathrsfs}
\newcommand{\R}{\mathbb{R}}
\newcommand{\E}{\operatorname{\mathbb{E}}}
\newcommand{\pdf}{\mathrm{pdf}}
\newcommand{\prob}{\mathrm{Prob}}
\newcommand{\sgn}{\operatorname{sgn}}
\newcommand{\abs}[1]{\left\vert#1\right\vert}
\renewcommand{\vec}[1]{\mathbf{#1}}
\def\ITM{\textit{ITM Probe}}
\def\la{\left\langle}
\def\ra{\right\rangle}
\newcommand{\refeq}[1]{(\ref{#1})}

\newcommand{\rev}[1]{#1}

\begin{document}

\begin{titlepage}

\begin{center}
{\Large\bf Robust and accurate data enrichment statistics via distribution function of sum of weights}
\end{center}
\vspace{.35cm}

\begin{center}
{\large Aleksandar Stojmirovi\'c\, and Yi-Kuo Yu\footnote{to whom correspondence should be addressed}}
\vspace{0.25cm}
\small

\par \vskip .2in \noindent
National Center for Biotechnology Information\\
National Library of Medicine\\
National Institutes of Health\\
Bethesda, MD 20894\\
United States
\end{center}

\normalsize
\vspace{0.25cm}

\begin{abstract}

\subsubsection*{Motivation:}
Term enrichment analysis facilitates biological interpretation by assigning to experimentally/computationally obtained data annotation associated with terms from controlled vocabularies. This process usually involves obtaining statistical significance for each vocabulary term and using the most significant terms to describe a given set of biological entities, often associated with weights. 
Many existing enrichment methods require selections of (arbitrary number of) the most significant entities and/or do not account for weights of entities. Others either mandate extensive simulations to obtain statistics \rev{or assume normal weight distribution}. In addition, most methods have difficulty assigning correct statistical significance to terms with few entities.

\subsubsection*{Results:}
Implementing the well-known Lugananni-Rice formula, we have developed a novel approach, called SaddleSum, that is free from all the aforementioned constraints and evaluated it against several existing methods. With entity weights properly taken into account, SaddleSum is internally consistent and stable with respect to the choice of number of most significant entities selected. Making few assumptions on the input data, the proposed method is universal and can thus be applied to areas beyond analysis of microarrays. Employing asymptotic approximation, SaddleSum  provides a term-size dependent score distribution function that gives rise to accurate statistical significance even for terms with few entities. As a consequence, SaddleSum enables researchers to place confidence in its significance assignments to small terms that are often biologically most specific.

\subsubsection*{Availability:}
Our implementation, which uses Bonferroni correction to account for multiple hypotheses testing, is available at \href{http://www.ncbi.nlm.nih.gov/CBBresearch/qmbp/mn/enrich/}{http://www.ncbi.nlm.nih.gov/CBBresearch/qmbp/mn/enrich/}. \rev{Source code for the standalone version can be downloaded from \href{ftp://ftp.ncbi.nlm.nih.gov/pub/qmbpmn/SaddleSum/}{ftp://ftp.ncbi.nlm.nih.gov/pub/qmbpmn/SaddleSum/}.}

\subsubsection*{Contact:} \href{yyu@ncbi.nlm.nih.gov}{yyu@ncbi.nlm.nih.gov}
\end{abstract}
\end{titlepage}

\section{Introduction}

A major challenge of contemporary biology is to ascribe interpretation to high-throughput experimental or computational results, where each considered entity (gene or protein) is assigned a value. Biological information is often summarized through controlled vocabularies such as Gene Ontology (GO) \citep{ABB00}, where each annotated term includes a list of entities. Let $\vec{w}$ denote a collection of values, each associated with an entity. Given $\vec{w}$ and a controlled vocabulary, enrichment analysis aims to retrieve the terms that by statistical inference best describe $\vec{w}$, that is, the terms associated with entities with atypical values. Many enrichment analysis tools have been developed primarily to process microarray data \citep{HSL09}. In terms of biological relevance, the performance assessment of those tools is generally difficult. It requires a large, comprehensive `gold standard' vocabulary together with a collection of $\vec{w}$'s processed from  experimental data, and with true/false positive terms corresponding to each $\vec{w}$ correctly specified. This invariably introduces some degree of circularity because the terms often come from curating experimental results. Before declaring efficacy in biological information retrieval that is nontrivial to assess, an enrichment method should pass at least the statistical accuracy and internal consistency test. 

In their recent survey, 
\cite{HSL09} list 68 distinct bioinformatic enrichment tools introduced between 2002 and 2008. Most tools share a similar workflow: given $\vec{w}$ obtained by suitably processing experimental data, they sequentially test each vocabulary term for enrichment to obtain its P-value (the likelihood of a false positive given the null hypothesis). Since many terms are tested, a multiple hypothesis correction, such as Bonferroni \citep{HT87} or false discovery rate (FDR) \citep{BH95}, is applied to P-value of each to obtain the final statistical significance. The results are displayed for the user in a suitable form outlining the significant terms and possibly relations between them. Note that the latter steps are largely independent from the first. To avoid confounding factors, we will focus exclusively on the original enrichment P-values.

Based on the statistical methods employed, the existing enrichment tools can generally be divided into two main classes. The singular enrichment analysis (SEA) class contains numerous tools that form the majority of published ones \citep{HSL09}. By ordering values in $\vec{w}$, these tools require users to select a number of top-ranking entities as input and mostly use hypergeometric distribution (or equivalently Fisher's exact test) to obtain the term P-values. After the selection is made, SEA treats all entities equally, ignoring their value differences. 

The gene set analysis (GSA) class was pioneered by the GSEA tool \citep{MLES03,STMM05}. Tools from this class use all values (entire $\vec{w}$) to calculate P-values and do not require pre-selection of entities. Some approaches \citep{BAH04,AADH07,BBVB07,ENSL09} in this group apply hypergeometric tests to all possible selections of top-ranking entities. The final P-value is computed by combining (in a tool-specific manner) the P-values from the individual tests. Other approaches use non-parametric approaches: rank-based statistics such as Wilcoxon rank-sum \citep{BEK04} or Kolmogorov-Smirnov-like \citep{MLES03,STMM05,BBS05,BKKK07}. When weights are taken into account, such as in GSEA \citep{STMM05}, statistical significance must be determined from a sampled (shuffled) distribution. Unfortunately, limited by the number of shuffles that can be performed, the smallest obtainable P-value is bounded away from 0. 

The final group of GSA methods computes a score for each vocabulary term as a sum of the values (henceforth used interchangeably with weights) of the $m$ entities it annotates. In general, the score distribution $\pdf_m(S)$ for the experimental data is unknown. By Central Limit Theorem, when $m$ is large, Gaussian \citep{SD04,KV05} or Student's t-distribution \citep{BFVK05,LFSH09} can be used to approximate $\pdf_m(S)$. Unfortunately, when the weight distributions are skewed, the required $m$ may be too large for practical use. Evidently, this undermines the P-value accuracy of small terms (meaning terms with few entities)\rev{, which are biologically most specific.}

It is generally found that, given the same vocabulary and $\vec{w}$, different enrichment analysis tools report diverse results. We believe this may be attributed to disagreement in P-values reported as well as that different methods have different degree of robustness (internal consistency). Instead of providing a coherent biological understanding, the array of diverse results questions the confidence of information found. Furthermore, other than microarray datasets, there exist experimental or computational results such as those from ChIP-chip \citep{ELYY07}, deep sequencing \citep{SSRM08}, quantitative proteomics \citep{SWBG09} and \textit{in silico} network simulations \citep{SY07,SY09}, that may benefit from enrichment analysis. It is thus imperative to have an enrichment method that report accurate P-values, preserves internal consistency, and allows investigations of a broader range of datasets.

To achieve these goals, we have developed a novel enrichment tool, called SaddleSum, that founds on the well-known Lugananni-Rice formula \citep{LR80} and derives its statistics from approximating asymptotically the distribution function of the scores used in the parametric GSA class. This allows us to obtain accurate statistics even in the cases where the distribution function generating $\vec{w}$ is very skewed and for terms containing few entities. The latter aspect is particularly important for obtaining biologically specific information. 

\section{Methods}

\begin{methods}
\subsection{Mathematical foundations for SaddleSum}
We distinguish two sets: the set of entities $\mathcal{N}$ of size $n$ and the controlled vocabulary $\mathcal{V}$. Each term from $\mathcal{V}$ maps to a set $\mathcal{M}\subset\mathcal{N}$ of size $m < n$. From experimental results, we obtain a 
set $\vec{w}=\{w_j\, |\, j\in\mathcal{N}\}$ and ask how likely it is to randomly pick $m$ entities whose sum of weights exceeds the sum $\hat{S}=\sum_{j\in \mathcal{M}} w_j$.

Assume that the weights in $\vec{w}$ come independently from a continuous probability space $W$ with the density function $p$ such that the moment generating function $\rho(t) = \int_W p(x) e^{t x} dx$ exists for $t$ in a neighborhood of 0. The density of $S$, sum of $m$ weights arbitrarily sampled from $\vec{w}$, can be expressed by the Fourier inversion formula 
\begin{align}
\pdf_m(S) &= \frac{1}{2\pi}\int_{-\infty}^\infty e^{m K(it)-it S}\ dt, \label{eqT:int3}
\end{align}
where $K(t)=\ln\rho(t)$ denotes the cumulant generating function of $p$. The tail probability or P-value for a score $\hat{S}$ is given by 
\begin{equation}\label{eqT:pval1}
\prob(S\geq \hat{S}) = \int_{\hat{S}}^\infty \pdf_m(S)\, dS.
\end{equation}
We propose to use an asymptotic approximation to \refeq{eqT:pval1}, which improves with increasing $m$ and $\hat{S}$.

\cite{Daniels54} derived an asymptotic approximation for the density $\pdf_m$ through saddlepoint expansion of the integral \refeq{eqT:int3} while the corresponding approximation to the tail probability was obtained by 
\cite{LR80}.  Let  $\phi(x)=\exp(-x^2/2)/\sqrt{2\pi}$ and $\Phi(x)=\int_x^\infty\phi(t)dt$ denote respectively the density and the tail probability of Gaussian distribution. Let $\hat{\lambda}$ be a solution of the equation
\begin{equation}\label{eqT:saddle1}
\hat{S} = mK'(\hat{\lambda}).
\end{equation}
Then, the leading term of the Lugananni-Rice approximation to the tail probability takes the form
\begin{equation}\label{eqT:tail}
\prob(S\geq \hat{S}) = \Phi(\hat{z}) + \left(\frac{1}{\hat{y}}- \frac{1}{\hat{z}} \right) \phi(\hat{z})+ O(m^{-3/2}),
\end{equation}
where $\hat{y}=\hat{\lambda}\sqrt{mK''(\hat{\lambda})}$ and $\hat{z}=\sgn(\hat{\lambda})\sqrt{2(\hat{\lambda}\hat{S}-mK(\hat{\lambda}))}$. Appropriate summary of derivation of \refeq{eqT:tail} is provided in Supplementary Materials.

\cite{Daniels54} has shown that eq.\,\refeq{eqT:saddle1} has a unique simple root under most conditions and that $\hat\lambda$ increases with $\hat{S}$, with $\hat\lambda=0$ for $\hat{S}=m \la W \ra$ where $\la W \ra = \int_W x p(x)\, dx$ is the mean of $W$. While the approximation \refeq{eqT:tail} is uniformly valid over the whole domain of $p$, its components need to be rearranged for numerical computation near the mean. When $\hat{S}\gg m \la W \ra$,  $\phi(\hat{z})/\hat{y}$ dominates and the overall error is $O(m^{-1})$ \citep{Daniels87}.

SaddleSum, our implementation of Lugananni-Rice approximation for computing enrichment P-values, first solves eq.\,\refeq{eqT:saddle1} for $\hat{\lambda}$ using Newton's method and then returns the P-value using \refeq{eqT:tail}. The derivatives of the cumulant generating function are estimated from $\vec{w}$: we approximate the moment generating function by \rev{$\rho(t)\approx \frac{1}{n}\sum_{j\in\mathcal{N}}e^{t w_j}$}, and then $K'(t)=\rho'(t)/\rho(t)$ and $K''(t)=\rho''(t)/\rho(t) - (K'(t))^2$. Since the same $\vec{w}$ is used to sequentially evaluate P-values of all terms in $\mathcal{V}$, we retain previously computed $\hat{\lambda}$ values in a sorted array. This allows us, using binary search, to reject many terms with P-values greater than a given threshold without running Newton's method or to bracket the root of \refeq{eqT:saddle1} for faster convergence. \rev{More details on the SaddleSum implementation and evaluations of its accuracy against some well-characterized distributions are in Section 2 of Supplementary Materials. When run as a term enrichment tool, SaddleSum reports E-value for each significant term by applying Bonferroni correction to the term's P-value.}

\subsection{Gene Ontology} The assignment of human genes to GO terms was taken from the NCBI gene2go file (ftp://ftp.ncbi.nih.gov/gene/DATA/gene2go.gz) downloaded on 07-02-2009. After assigning all genes to terms, we removed all redundant terms -- if several terms mapped to the same set of genes, we kept only one such term. For our statistical experiments we kept only the terms with no less than five mapped genes within the set of weights considered and hence the number of processed terms varied for each realization of sampling (see below).

\subsection{Information flow in protein networks}
\ITM\ \citep{SY09} is an implementation of the framework for exploring information flow in interaction networks \citep{SY07}. Information flow is modeled through discrete time random walks with damping -- at each step the walker has a certain probability of leaving the network. Although \ITM\ offers three modes: emitting, absorbing and channel, we only used the simplest, emitting mode, to provide examples illustrating issues of significance assignment. The emitting mode takes as input one or more network proteins, called sources, and a damping factor $\alpha$. For each protein node in the network, the model outputs the expected number of visits to that node by random walks originating from the sources, thus highlighting the network neighborhoods of the sources. The damping factor determines the average number of steps taken by a random walk before termination: $\alpha=1$ corresponds to no termination while $\alpha=0$ leads to no visits apart from the originating node. For our protein-protein interaction network examples, we used the set of all human physical interactions from the BioGRID \citep{BSRB08}, version 2.0.54 (July 2009). The network consists of 7702 proteins and 56400 unique interactions. Each interaction was represented by an undirected link. A link carries weight 2 if its two ends connect to the same protein and 1 otherwise. 

\subsection{Microarrays} From the NCBI Gene Expression Omnibus (GEO) \citep{BTWL09},
we retrieved human microarray datasets with expression $\log_2$ ratios (weights) provided, resulting in 34 datasets and 136 samples in total. For each sample, when multiple weights for the same gene were present, we took their mean instead. This resulted in a $\vec{w}$  where each gene is assigned a unique raw weight. For evaluations, we also used another version of $\vec{w}$  where negative weights were set to zero. This version facilitated investigation of up-regulation while keeping the down-regulated genes as part of statistical background. 

\subsection{Evaluating accuracy of P-values} By definition, a P-value associated with a score is the probability of that score or better arising purely by chance. We tested the accuracy of reported P-values reported by enrichment methods via simulations on `decoy' databases, which contained only terms with random gene assignments. For each term from the decoy dataset and each set of weights based on network or microarray data, we recorded the reported P-value and thus built an empirical distribution of P-values. If a method reports accurate P-values, the proportion of runs, which we term empirical P-value, reporting P-values smaller than or equal to a P-value cutoff, should be very close to that cutoff. We show the results graphically by plotting on the log-log scale the empirical P-value as a function of the cutoff.

For each given list of entities $\mathcal{N}$, be it from the target gene set of a microarray dataset or the set of participating human proteins in the interaction network, we produced two types of decoy databases. The first type was based on GO. We shuffled gene labels 1000 times. For each shuffle, we associated all terms from GO with the shuffled labels to retain the term dependency. This resulted in a database with approximately \rev{$5\times 10^6$} terms (1000 shuffles times about \rev{5000} GO terms). In the second type, each term, having the same size $m$, was obtained by sampling without replacement $m$ genes from $\mathcal{N}$. The databases from this type (one for each term size considered) contained exactly $10^7$ terms. The evaluation query set of 100 $\vec{w}$'s from interaction networks was obtained by randomly sampling 100 proteins out of 7702 and running \ITM\ with each protein as a single source. The weights for source proteins were not considered since they were prescribed, not resulting from simulation. Each run used $\alpha=0.7$, without excluding any nodes from the network. For microarrays, the set of 136 samples was used. Since both query sets are of size $\approx\!\!10^2$, the total number of $\vec{w}$--term matches was $\approx\!\!10^9$.

\subsection{Student's t-test (used by GAGE and T-profiler)} Similar to SaddleSum, t-test approaches are based on sum-of-weights score but use the Student's t-distribution to infer P-values. As before, let $w_j$ denote the weight associated with entity $j\in\mathcal{N}$, let $\mathcal{M}$ denote the set of $m$ entities associated with a term from vocabulary and let $\mathcal{M}'=\mathcal{N}\setminus\mathcal{M}$. For any set $\mathcal{S}\subseteq\mathcal{N}$ of size $\abs{\mathcal{S}}$, let $x_\mathcal{S}=\frac{1}{\abs{\mathcal{S}}}\sum_{j\in\mathcal{S}} w_j$ denote the mean weight of entities in $\mathcal{S}$ and let $s^2_\mathcal{S}=\frac{1}{\abs{\mathcal{S}}-1}\sum_{j\in\mathcal{S}} w_j-x_\mathcal{S})^2$ be their sample variance.

GAGE \citep{LFSH09} enrichment tool uses two sample t-test assuming unequal variances and equal sample sizes to compare the means over $\mathcal{N}$ and $\mathcal{M}$. The test statistic is
\begin{equation}\label{eq:ttst2}
t = \frac{x_\mathcal{M}-x_\mathcal{N}}{\sqrt{s^2_\mathcal{M}/m + s^2_\mathcal{N}/m}}
\end{equation}
and the P-value is obtained from the upper tail of the Student's t-distribution with degrees of freedom
\begin{equation*}
\nu = (m-1)\frac{(s^2_\mathcal{M}+s^2_\mathcal{N})^2}{s^4_\mathcal{M}+s^4_\mathcal{N}}.
\end{equation*}
T-profiler \citep{BFVK05} compares the means over $\mathcal{M}$ and $\mathcal{M}'$ using two sample t-test assuming equal variances but unequal sample sizes. The pooled variance estimate is given by
\begin{equation*}
s^2 = \frac{(m-1)s^2_\mathcal{M} + (n-m-1)s^2_{\mathcal{M}'}}{n-2},
\end{equation*}
and the test statistic is 
\begin{equation*}
t = \frac{x_\mathcal{M}-x_{\mathcal{M}'}}{s\sqrt{\frac{1}{m}+\frac{1}{n-m}}}.
\end{equation*}
The T-profiler P-value is then obtained from the tail of the Student's t-distribution with $\nu=n-2$ degrees of freedom.

\subsection{Hypergeometric distribution} Methods based on hypergeometric distribution or equivalently, Fisher's exact test, use only rankings of weights and require selection of `significant' entities prior to calculation of P-value. We first rank all entities according to their weights and consider the set $\mathcal{C}$ of $c$ entities with largest weights. The number $c$ can be fixed (say 50), correspond to a fixed percentage of the total number of weights, depend on the values of weights, or be calculated by other means. The score $\hat{S}$ for the term $\mathcal{M}$ is given by the size of the intersection, $\mathcal{C}\cap\mathcal{M}$, between $\mathcal{C}$ and $\mathcal{M}$. This is equivalent to setting $\hat{S}=\sum_{j\in \mathcal{M}} w_j$ with $w_j=1$ for $j\in\mathcal{C}$ and 0 otherwise. The P-value for score $\hat{S}$ is
\begin{equation*} 
\rev{\prob(S\geq\hat{S}) = \sum_{i=\hat{S}}^{\min(c,m)} \frac{\binom{m}{i}\binom{n-m}{c-i}}{\binom{n}{c}}.}
\end{equation*}
Hence, the P-value measures the likelihood of score $\hat{S}$ or better over all possible ways of selecting $c$ entities out of $\mathcal{N}$, with $m$ entities associated with the term investigated.

In each of our P-value accuracy experiments we used two variants of the hypergeometric method, one taking a fixed percentage of nodes and the other taking into account the values of weights. For microarray datasets, the fist variant took 1\% of available genes (HGEM-PN1) while the second select genes with four fold change or more (HGEM-F2). In experiments based on protein networks, we took 3\% of available proteins (231 entities) for the first variant (HGEM-PN3) and used the participation ratio formula to determine $c$ in the second (HGEM-PR). Participation ratio \citep{SY07} is given by the formula
\begin{equation*}
c = \frac{\left(\sum_{i\in \mathcal{N}} w_i \right)^2}{\sum_{j\in \mathcal{N}} w_j^2}.
\end{equation*}
We chose a smaller percentage of weights for microarray-based data (1\% vs 3\% for data derived for networks) because the microarray datasets generally contained measurement for more genes than the number of proteins in the network. 

\begin{figure*}
\includegraphics{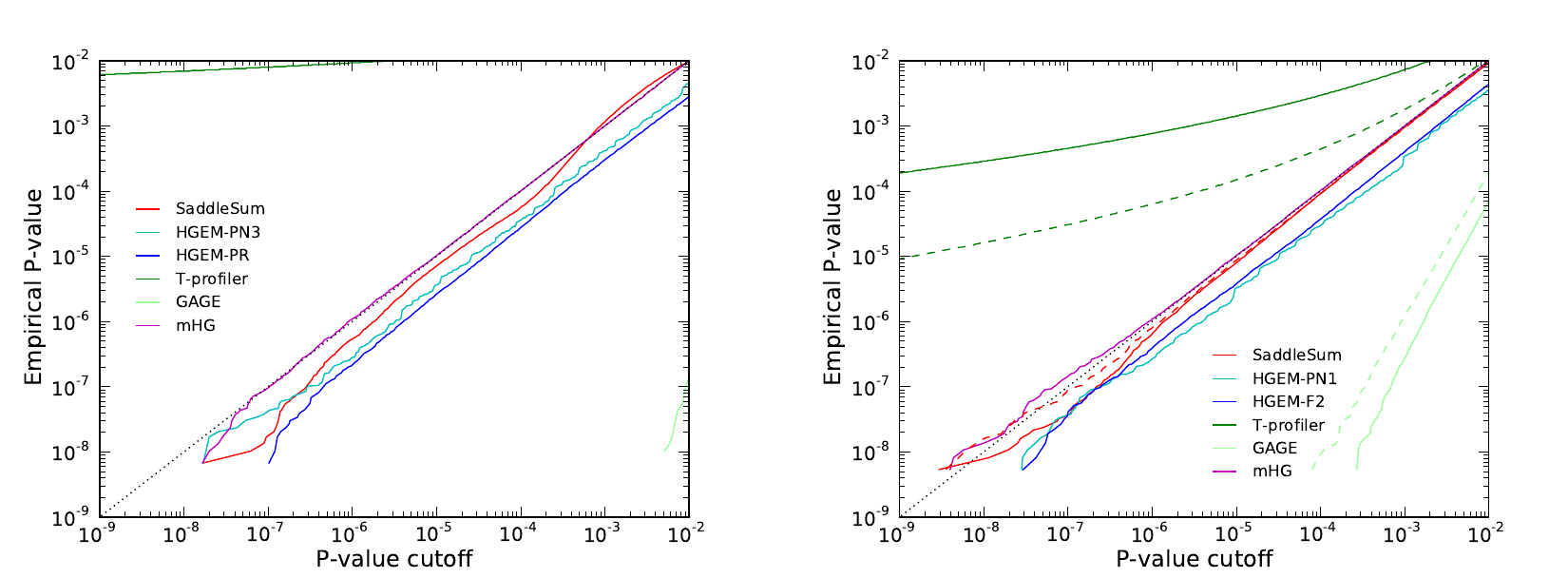}
\caption{Empirical P-values versus P-value cutoffs reported for investigated enrichment methods. Methods with accurate statistics have their curves follow the dotted line closely over the whole range. Each curve was constructed by aggregating the results of approximately $10^9$ GO-based decoy term queries. Displayed on the left (right) are results using weights derived from protein network information flow simulations (microarrays). In microarray plots for SaddleSum, T-profiler and GAGE, full lines indicate the results where negative weights were set to 0, while dashed lines show the results using all weights. The reason  that HGEM curves run below the theoretical line and parallel to it is that every curve is an aggregate of many curves, each of which (i) represents a single sample of weights determining parameters to be fed into hypergeometric distribution, and (ii) is a step function touching the theoretical line and dropping below it. Merging curves from many samples produces the effect seen in our plots.}\label{fig:stats1} 
\end{figure*}

\subsection{mHG score} Instead of making a single, arbitrary choice of $c$ and applying hypergeometric score, mHG method implemented in the GOrilla package \citep{ENSL09} considers all possible $c$'s. The mHG score is defined as
\begin{equation*}
\rev{mHG = \min_{1\leq c\leq n}\sum_{i=k}^{\min(c,m)} \frac{\binom{m}{i}\binom{n-m}{c-i}}{\binom{n}{c}},}
\end{equation*}
where $k$ is the number of entities annotated by the term $\mathcal{M}$ among the $c$ top-ranked entities. The exact P-value for mHG score is then calculated by using a dynamic programming algorithm developed by Eden \textit{et al.} \citep{ELYY07}. For our experiments we used an implementation in C programming language that was derived from the Java implementation used by GOrilla. The implementation uses a truncated algorithm that gives an approximate P-value with improved running speed.

\subsection{Retrieval stability with respect to choice of $\mathbf{\emph{c}}$} To evaluate consistency of investigated methods, we compared the sets of significant terms retrieved from GO using different numbers of nonzero weights as input. For each $\vec{w}$, we sort in descending order the weights associated with entities. With each $c$ selected, we  kept $c$ largest weights unchanged and set the remaining to 0 to arrive at a modified set of weights $\vec{w}\vert\mathcal{C}$. We did not totally exclude the lower weights but kept them under consideration to provide statistical background. We submitted $\vec{w}\vert\mathcal{C}$ for analysis and obtained from each statistical method a set of enriched terms ordered by their P-value. In Fig.\,\ref{fig:mainexamples1}A and Supplementary Fig.\,S3, we displayed the actual five most significant terms retrieved with their P-values for selected examples of weight sets. To investigate on a larger scale the retrieval stability to $c$ changes, we computed for each method the overlap between sets of top ten terms from two different $c$'s for the $\vec{w}$ sets mentioned in `Evaluating accuracy of P-values' and then took the average (Fig.\,\ref{fig:mainexamples1}B).
\end{methods}

\begin{figure*}
\begin{center}
\includegraphics[scale=0.925]{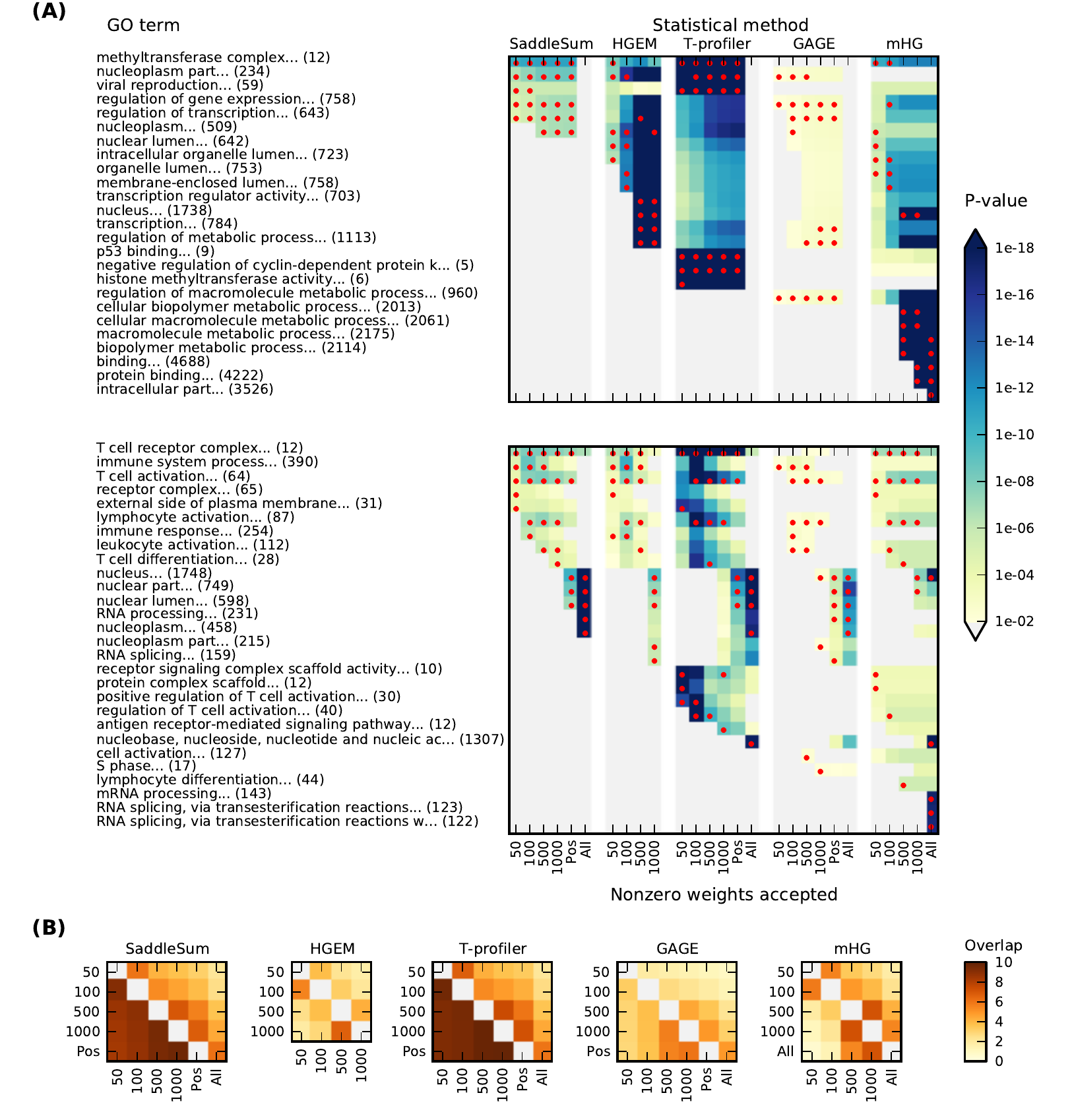}
\caption{P-value consistency and retrieval stability. \textbf{(A)} The output of \ITM\ emitting mode with human MLL protein (histone methyltransferase subunit) as the source (top) and the $\log_2$ ratios from the human T cell signaling microarray GSM89756 (bottom) were processed by each of the five investigated statistical methods with varying number of weighted entities included for analysis (\textsl{All} and \textsl{Pos} include all entities; \textsl{All} uses raw weights while \textsl{Pos} sets all negative weights to 0). The P-values for GO terms from the union of the sets of top-five hits for each method and different numbers of selected entities, are indicated by colors of the corresponding cell. Red dots show the actual top five hits for the method represented by that column. \textbf{(B)} Degree of overlap between sets of significant GO terms. Each panel corresponds to a single method with different numbers of entities used for analysis, with the results from microarray queries shown in the upper triangle and those based on network flow shown in the lower triangle. Color in each cell indicates the average pairwise overlap between the two sets of top-ten entities retrieved. For example, consider the light orange colored cell (horizontally labeled by 100 and vertically labeled by 500) in the mHG panel. This indicates that on average the top-ten terms retrieved by mHG using top 100 and top 500 network flow proteins share about three common terms.}\label{fig:mainexamples1}
\end{center}
\end{figure*}

\section{Results}
We compared our SaddleSum approach against the following existing methods: 
Fisher's exact test (HGEM \citep{BWGJBCS04}), two sample Student's t-test with equal (T-profiler \citep{BFVK05}) and unequal (GAGE \citep{LFSH09}) variances, and mHG score \citep{ELYY07,ENSL09}. Based on data from both microarrays and simulations of information flow in protein networks, the comparison shown here encompassed (in order of importance) evaluation of P-value accuracy, ranking stability and running time. Accurate P-value reflects the likelihood of a false identification and thus allows for comparison between terms retrieved even across experiments. Incorrect P-values therefore render ranking stability and algorithmic speed pointless. Accurate P-values without ranking stability question the robustness of biological interpretation. For pragmatic use of an enrichment method, even with accurate statistics and stability, it is still important to have reasonable speed.

\subsection{Accuracy of reported P-values}
The term P-value reported by an enrichment analysis method provides the likelihood for that term to be enriched within $\vec{w}$. To infer biological significance using statistical analysis, it is essential to have accurate P-values. We analyzed the accuracy of P-values reported by the investigated approaches through simulating $\approx\!\!10^9$ queries and comparing their reported and empirical P-values. 

Results based on querying databases with fixed term sizes are shown in Supplementary Figs.\,S1 and S2. 
Shown in Fig.\,\ref{fig:stats1} are the results for querying GO-based gene-shuffled term databases, which retain the structure of the original GO as a mixture of terms of different sizes organized as a directed acyclic graph where small terms are included in larger ones. The curves for all methods in Fig.\,\ref{fig:stats1} therefore resemble a mixture of curves from Supplementary Figs.\,S1 and S2 albeit weighted towards smaller-sized terms.

For weights from both network simulations and microarrays, SaddleSum as well as the methods based on Fisher's exact test (HGEM and mHG) report P-values that are acceptable (within one order of magnitude from the theoretical values). For HGEM and mHG, this is not surprising because our experiments involved shuffling entity labels and hence followed the null model of the hypergeometric distribution. On the other hand, the null model of SaddleSum and the t-test methods assumes weights drawn independently from some distribution (sampling with replacement). For terms with few entities \rev{($m \leq 100$)}, the difference between the two null models is minimal and the P-value accuracy assessment curves for SaddleSum run as \rev{close} to the theoretical line \rev{as those for HGEM methods}. For \rev{$m > 100$}, SaddleSum gives more conservative P-values for terms with large sums of weights (Supplementary Figs.\,S1 and S2). In practice, this has no significant effect to biological inference. Large terms would be still selected as significant given a reasonable P-value cutoff and accurate P-values are assigned to small terms that are biologically specific.

Two-sample t-test with unequal variances as used by GAGE package reports P-values so conservative that they are often larger than 0.01 and hence not always visible in our accuracy plots. This effect persists even for $m$ as large as 500. This might be because the number of degrees of freedom used is considerably small. In addition, its test statistic (eq.\,\refeq{eq:ttst2}) emphasizes the estimated within-term variance $s^2_\mathcal{M}$ that is typically larger than the overall variance $s^2_\mathcal{N}$. 

On the other hand, T-profiler generally exaggerates \citep{LFSH09} significance because it uses the t-distribution with a large number of degrees of freedom ($n-2$).  Although some small terms may appear biologically relevant (as in Fig.\,\ref{fig:mainexamples1}), one should not equate these exaggerated P-values with sensitivity. For microarray data, the $\log_2$ ratios are almost symmetrically distributed about 0 (Supplementary Fig.\,S4). The distribution of their sum is close to Gaussian. However, T-profiler still significantly exaggerates P-values for terms whose $m<25$ (Supplementary Fig.\,S2). The statistical accuracy of T-profiler worsens when negative $\log_2$ ratios are set to 0. The reason for doing so is that allowing weights within each term to cancel each other may not be  biologically appropriate. GO terms may cover a very general category where annotations may not always be available for more specific subterms. Subsequently, terms may get refined and new terms may emerge. In such situation, it is desirable to discover terms that have genes that are significantly up-regulated even if many genes from the same term are down-regulated. 

\subsection{Stability}
P-value accuracy, although the most important criterion, measures only performance with respect to non-significant hits, that is, the likelihood of a false positive. It is also necessary to consider the quality of enrichment results in terms of the underlying biology. Testing the quality directly, as described in the introduction, is not yet feasible. Instead we evaluated internal consistency of each method with respect to the number of top-ranked entities used for analysis. Fig.\,\ref{fig:mainexamples1}A shows the change of P-values reported for the top five GO terms with respect to the number of selected entities using two examples with weights respectively from network flow simulation and microarray. Additional examples are shown in Supplementary Fig.\,S3. Results from evaluating the overall consistency of the best ten terms retrieved are shown in Fig.\,\ref{fig:mainexamples1}B.

Both HGEM and mHG methods are highly sensitive to the choice of $c$, the number of entities deemed significant. With a small $c$, their sets of significant terms resemble the top terms obtained by SaddleSum, while large values of $c$ render very small P-values for large-sized terms (often biologically non-specific). This is mainly because HGEM and mHG treat all selected significant entities as equally important without weighting down less significant entities, the collection of which may out vote the most significant ones. Hence, although mHG considers all possible $c$ values, to obtain biologically specific interpretation, it might be necessary to either remove very large terms from the vocabulary or to impose an upper bound on $c$. In that respect, mHG is very similar to the original GSEA method \citep{MLES03}, which also ignored weights. The authors of GSEA noted that the genes ranked in middle of the list had disproportionate effect to their results and produced an improved version of GSEA \citep{STMM05} with weights considered.

GAGE does not show strong consistency because many P-values it reports are too conservative and fall above the 0.01 threshold we used. Consequently, the best overlap between various cutoffs is about 5 (out of 10) for network flow examples and 4 for microarray examples (Fig.\,\ref{fig:mainexamples1}B). T-profiler shows great internal consistency. Unfortunately, as shown in Fig.\,\ref{fig:stats1}, Supplementary Figs.\,S1 and S2, it reports inaccurate P-values, especially for small terms. This is illustrated in the top panel of Fig.\,\ref{fig:mainexamples1}A, where T-profiler selects as highly significant the small terms (with 5,6 and 9 entities), which are deemed insignificant by all other methods. The same pattern can be observed in Supplementary Fig.\,S3, although the severity is tamed for microarrays. Using weights for scoring terms, SaddleSum is also stable with respect to the choice of $c$ but with accurate statistics.

\subsection{Speed}
In terms of algorithmic running time \rev{(Table~\ref{tbl:speed})}, parametric methods relying on normal or Student's t-distribution require few computations. Methods based on hypergeometric distribution, if properly implemented, are also fast. On the other hand, non-parametric methods can take significant time if many shufflings are performed. Based on dynamic programming, mHG method can also take excessive time for large terms. SaddleSum has running time that is only slightly longer than that of parametric methods. 

\begin{table}
\rev{\footnotesize
\begin{tabular}{@{\vrule height 10.5pt depth4pt  width0pt}lrr@{\hspace{5mm}}rr} \vspace{-0mm}
& \multicolumn{2}{l}{Total running time} & \multicolumn{2}{l}{Average time per query} \\ 
Method & network & microarray & network & microarray \\ \hline
SaddleSum & 558 & 872 & 0.56 & 0.64 \\
HGEM & 501 & 615 & 0.50 & 0.45 \\
T-profiler & 446 & 586 & 0.45 & 0.43 \\
GAGE & 499 & 651 & 0.50 & 0.48 \\
mHG & 2433 & 3407 & 2.43 & 2.51 \\ \hline \\
\end{tabular}
\caption{Running times of evaluated enrichment statistics algorithms (in seconds). We queried GO ten times with each of the five examined enrichment methods using weights from 100 network simulation results and 136 microarrays (same datasets used for P-value accuracy experiments). Running times for P-value calculations on dual-core 2.8 GHz AMD Opteron 254 processors (using a single core for each run) aggregated over all samples are shown on the left, while average times per query are shown on the right. The HGEM method used 100-object cutoff.}\label{tbl:speed}}
\end{table}

\section{Discussion}

Approximating the distribution of sum of weights by saddlepoint method, our SaddleSum is able to adapt itself equally well to distributions with widely different properties. The reported P-values have accuracy comparable to that of the methods based on the hypergeometric distribution while requiring no prior selection of the number of significant entities.

While our results show that GAGE method suffers from reduced sensitivity, it should be noted that it forms only a part of GAGE algorithm. GAGE was designed to compare two groups of microarrays (for example disease and control) by obtaining an overall P-value. In that scheme, the P-values we evaluated are used only for one-on-one comparisons between members of two groups. By combining one-on-one P-values (which are assumed independent), the overall P-value obtained by GAGE can become quite small. 

The assumed null distribution by T-profiler \citep{BFVK05} is close to Gaussian. It has been commented \citep{LFSH09} that its statistics are similar to that of PAGE \citep{KV05}, which uses Z-test. Naturally, the smallest, and likely exaggerated, P-values occur when evaluating small terms. For that reason, PAGE does not consider terms with less than 10 entities, which we included in our evaluation solely for the purpose of comparison.

Our network simulation experiments produce very different weight profiles (Supplementary Fig.\,S4) than that of microarrays. These weights are always positive and skewedly distributed. Even after summing many such weights, the distribution of the sum is still far from Gaussian in the tail. Therefore, T-profiler and GAGE are unable to give accurate statistics.  Overall, our evaluations clearly illustrate the inadequacy, even for large terms, of assuming nearly Gaussian null distribution when the data is skewed. While Central Limit Theorem does guarantee convergence to Gaussian for large $m$, the convergence may not be sufficiently fast in the tail regions, which influence the statistical accuracy the most.

As presented here, SaddleSum uses given $\vec{w}$ both for estimating the $m$-dependent score distribution and for scoring each term. If a certain distribution of weights are prescribed, it is possible to adapt our algorithm to take a histogram for that distribution as input and use experimentally obtained weights for scoring only. 

A possible way to improve biological relevance in retrieval is to allow for term-specific weight assignment. For example, a gene associated with a GO term can be assigned a `NOT' qualifier to indicate explicitly  that this gene product is not associated with the term considered. A way to use this information would be to change the sign of the weight for such a gene (from positive to negative or vice versa), but only when scoring the terms where the qualifier applies. Hence, potentially every term could be associated with a specific weight distribution. While all methods using weights can implement this scheme, SaddleSum is particularly suitable for it because it handles well the small terms and skewed distributions, where changing the sign for a single weight can have a considerable effect. This procedure can be generalized so that each gene in a term carries a different weight.

\rev{Several authors \citep{HSL09,GB07,GCWM07} have raised the issue of correlation between
 weights of entities: generally the weights of biologically related genes or proteins 
 change together and therefore a null model assuming independence between weights may 
 result in exaggerated P-values. In principle, a good null model is one that can bring 
 out the difference between signal and noise. To what level of sophistication a null model
  should be usually is a trade-off between statistical accuracy and retrieval sensitivity. 
 Using protein sequence comparison for example, ungapped alignment enjoys a theoretically characterizable 
  statistics \citep{KA_90} but is not as sensitive as the gapped alignment \citep{PSI_BLAST}, where the
 score statistics is known
  only empirically because the null model allows for insertions and deletions of amino acids.  
 Incorporating  insertion and deletion into the null model made all the difference in 
  retrieval sensitivity. This is probably because insertions/deletions do occur abundantly
   in natural evolution of protein sequences. The ignorance of protein sequence correlations, 
   assumed by both ungapped and gapped alignments, does not seem to cause much harm in retrieval efficacy.}   
    
\rev{Although SaddleSum assumes weight independence and thus bears the possibility of exaggerating 
 statistical significance of an identified term, it mitigates this issue by incorporating 
 the entire $\vec{w}$ in the null distribution. It includes the entities with extreme 
 weights that clearly represent `signal' and not `noise', bringing higher the tail of the score 
 distribution and thus larger P-values. Indeed, as shown by examples in Fig.\,\ref{fig:mainexamples1}A and 
 Supplementary Fig.\,S3, SaddleSum does not show unreasonably small P-values. It should also be noted 
 that SaddleSum is designed for the simple case where a summary value is available for each entity considered
  -- its use for analyzing complex microarray experiments with many subjects divided into several groups 
  is beyond the scope of this paper and care must be exercised when using it in this context.}

SaddleSum is a versatile enrichment analysis method. Researchers are free to process appropriately their experimental data, produce a suitable $\vec{w}$ as input, and receive accurate term statistics from SaddleSum. Since it does not make many assumptions about the distribution of data, we foresee a number of additional applications not limited to genomics or proteomics, for example to literature searches.

\section*{Acknowledgments}
This work was supported by the Intramural Research Program of the National Library of Medicine at National Institutes of Health. This study utilized the high-performance computational capabilities of the Biowulf Linux cluster at the National Institutes of Health, Bethesda, MD. (http://biowulf.nih.gov).
We thank John Wootton and David Landsman for useful comments, Roy Navon for providing us with the Java source code for the statistical algorithms of GOrilla, Weijun Luo for his help with using GAGE package \rev{and the anonymous referees for comments that helped improve the first version of this paper.}


\begin{thebibliography}{}

\bibitem[Al-Shahrour {\em et~al.}(2007)Al-Shahrour, Arbiza, Dopazo,
  Huerta-Cepas, M{\'i}nguez, Montaner, and Dopazo]{AADH07}
Al-Shahrour, F. {\em et~al.} (2007).
\newblock From genes to functional classes in the study of biological systems.
\newblock {\em BMC Bioinformatics\/}, {\bf 8}, 114.

\bibitem[Altschul {\em et~al.}(1997)Altschul, Madden, Sch\"affer, Zhang, Zhang,
  Miller, and Lipman]{PSI_BLAST}
Altschul, S.~F. {\em et~al.} (1997).
\newblock Gapped {BLAST} and {PSI-BLAST}: a new generation of protein database
  search programs.
\newblock {\em Nucleic Acids Res.}, {\bf 25}, 3389--3402.

\bibitem[Ashburner {\em et~al.}(2000)Ashburner, Ball, Blake, Botstein, Butler,
  Cherry, Davis, Dolinski, Dwight, Eppig, Harris, Hill, Issel-Tarver,
  Kasarskis, Lewis, Matese, Richardson, Ringwald, Rubin, and Sherlock]{ABB00}
Ashburner, M. {\em et~al.} (2000).
\newblock Gene ontology: tool for the unification of biology. the gene ontology
  consortium.
\newblock {\em Nat Genet\/}, {\bf 25}, 25--29.

\bibitem[Backes {\em et~al.}(2007)Backes, Keller, Kuentzer, Kneissl, Comtesse,
  Elnakady, M{\"u}ller, Meese, and Lenhof]{BKKK07}
Backes, C. {\em et~al.} (2007).
\newblock {GeneTrail}--advanced gene set enrichment analysis.
\newblock {\em Nucleic Acids Res\/}, {\bf 35}(Web Server issue), W186--192.

\bibitem[Barrett {\em et~al.}(2009)Barrett, Troup, Wilhite, Ledoux, Rudnev,
  Evangelista, Kim, Soboleva, Tomashevsky, Marshall, Phillippy, Sherman,
  Muertter, and Edgar]{BTWL09}
Barrett, T. {\em et~al.} (2009).
\newblock {NCBI GEO: archive for high-throughput functional genomic data}.
\newblock {\em Nucleic Acids Res\/}, {\bf 37}(Database issue), D885--890.

\bibitem[Ben-Shaul {\em et~al.}(2005)Ben-Shaul, Bergman, and Soreq]{BBS05}
Ben-Shaul, Y. {\em et~al.} (2005).
\newblock Identifying subtle interrelated changes in functional gene categories
  using continuous measures of gene expression.
\newblock {\em Bioinformatics\/}, {\bf 21}(7), 1129--1137.

\bibitem[Benjamini and Hochberg(1995)Benjamini and Hochberg]{BH95}
Benjamini, Y. and Hochberg, Y. (1995).
\newblock Controlling the false discovery rate: a practical and powerful
  approach to multiple testing.
\newblock {\em Journal of the Royal Statistical Society\/}, {\bf 57}, 289--300.

\bibitem[Bleistein(1966)Bleistein]{Bleistein66}
Bleistein, N. (1966).
\newblock Uniform asymptotic expansions of integrals with stationary points and
  algebraic singularity.
\newblock {\em Communications in Pure and Applied Mathematics\/}, {\bf 19},
  353--370.

\bibitem[Blom {\em et~al.}(2007)Blom, Bosman, {van Hijum}, Breitling, Tijsma,
  Silvis, Roerdink, and Kuipers]{BBVB07}
Blom, E.-J. {\em et~al.} (2007).
\newblock {FIVA}: Functional information viewer and analyzer extracting
  biological knowledge from transcriptome data of prokaryotes.
\newblock {\em Bioinformatics\/}, {\bf 23}(9), 1161--1163.

\bibitem[Boorsma {\em et~al.}(2005)Boorsma, Foat, Vis, Klis, and
  Bussemaker]{BFVK05}
Boorsma, A. {\em et~al.} (2005).
\newblock {T-profiler}: scoring the activity of predefined groups of genes
  using gene expression data.
\newblock {\em Nucleic Acids Res\/}, {\bf 33}(Web Server issue), W592--595.

\bibitem[Boyle {\em et~al.}(2004)Boyle, Weng, Gollub, Jin, Botstein, Cherry,
  and Sherlock]{BWGJBCS04}
Boyle, E.~I. {\em et~al.} (2004).
\newblock {GO::TermFinder}--open source software for accessing gene ontology
  information and finding significantly enriched gene ontology terms associated
  with a list of genes.
\newblock {\em Bioinformatics\/}, {\bf 20}, 3710--3715.

\bibitem[Breitkreutz {\em et~al.}(2008)Breitkreutz, Stark, Reguly, Boucher,
  Breitkreutz, Livstone, Oughtred, Lackner, B{\"a}hler, Wood, Dolinski, and
  Tyers]{BSRB08}
Breitkreutz, B. {\em et~al.} (2008).
\newblock {The BioGRID Interaction Database: 2008 update}.
\newblock {\em Nucleic Acids Res\/}, {\bf 36}(Database issue), D637--640.

\bibitem[Breitling {\em et~al.}(2004)Breitling, Amtmann, and Herzyk]{BAH04}
Breitling, R. {\em et~al.} (2004).
\newblock Iterative group analysis ({iGA}): a simple tool to enhance
  sensitivity and facilitate interpretation of microarray experiments.
\newblock {\em BMC Bioinformatics\/}, {\bf 5}, 34.

\bibitem[Breslin {\em et~al.}(2004)Breslin, Ed{\'e}n, and Krogh]{BEK04}
Breslin, T. {\em et~al.} (2004).
\newblock Comparing functional annotation analyses with {Catmap}.
\newblock {\em BMC Bioinformatics\/}, {\bf 5}, 193.

\bibitem[Daniels(1954)Daniels]{Daniels54}
Daniels, H.~E. (1954).
\newblock Saddlepoint approximations in statistics.
\newblock {\em Ann. Math. Statist.}, {\bf 25}, 631--650.

\bibitem[Daniels(1987)Daniels]{Daniels87}
Daniels, H.~E. (1987).
\newblock Tail probability approximations.
\newblock {\em Internat. Statist. Rev.}, {\bf 55}(1), 37--48.

\bibitem[Eden {\em et~al.}(2007)Eden, Lipson, Yogev, and Yakhini]{ELYY07}
Eden, E. {\em et~al.} (2007).
\newblock Discovering motifs in ranked lists of dna sequences.
\newblock {\em PLoS Comput Biol\/}, {\bf 3}(3), e39.

\bibitem[Eden {\em et~al.}(2009)Eden, Navon, Steinfeld, Lipson, and
  Yakhini]{ENSL09}
Eden, E. {\em et~al.} (2009).
\newblock {GOrilla}: a tool for discovery and visualization of enriched go
  terms in ranked gene lists.
\newblock {\em BMC Bioinformatics\/}, {\bf 10}, 48.

\bibitem[Goeman and B\"uhlmann(2007)Goeman and B\"uhlmann]{GB07}
Goeman, J. and B\"uhlmann, P. (2007).
\newblock {Analyzing gene expression data in terms of gene sets: methodological
  issues}.
\newblock {\em Bioinformatics\/}, {\bf 23}(8), 980--987.

\bibitem[Gold {\em et~al.}(2007)Gold, Coombes, Wang, and Mallick]{GCWM07}
Gold, D. {\em et~al.} (2007).
\newblock {Enrichment analysis in high-throughput genomics - accounting for
  dependency in the NULL}.
\newblock {\em Brief Bioinform\/}, {\bf 8}(2), 71--77.

\bibitem[Hochberg and Tamhane(1987)Hochberg and Tamhane]{HT87}
Hochberg, Y. and Tamhane, A.~C. (1987).
\newblock {\em Multiple Comparison Procedures (Wiley Series in Probability and
  Statistics)\/}.
\newblock Wiley.

\bibitem[Huang {\em et~al.}(2009)Huang, Sherman, and Lempicki]{HSL09}
Huang, D.~W. {\em et~al.} (2009).
\newblock Bioinformatics enrichment tools: paths toward the comprehensive
  functional analysis of large gene lists.
\newblock {\em Nucleic Acids Res\/}, {\bf 37}(1), 1--13.

\bibitem[Jensen(1995)Jensen]{Jensen95}
Jensen, J.~L. (1995).
\newblock {\em Saddlepoint approximations\/}.
\newblock Clarendon Press, Oxford.

\bibitem[Karlin and Altschul(1990)Karlin and Altschul]{KA_90}
Karlin, S. and Altschul, S.~F. (1990).
\newblock Methods for assessing the statistical significance of molecular
  sequence features by using general scoring schemes.
\newblock {\em Proc. Natl. Acad. Sci. USA\/}, {\bf 87}, 2264--2268.

\bibitem[Kim and Volsky(2005)Kim and Volsky]{KV05}
Kim, S.-Y. and Volsky, D.~J. (2005).
\newblock {PAGE}: parametric analysis of gene set enrichment.
\newblock {\em BMC Bioinformatics\/}, {\bf 6}, 144.

\bibitem[Lugannani and Rice(1980)Lugannani and Rice]{LR80}
Lugannani, R. and Rice, S. (1980).
\newblock Saddle point approximation for the distribution of the sum of
  independent random variables.
\newblock {\em Adv. in Appl. Probab.}, {\bf 12}(2), 475--490.

\bibitem[Luo {\em et~al.}(2009)Luo, Friedman, Shedden, Hankenson, and
  Woolf]{LFSH09}
Luo, W. {\em et~al.} (2009).
\newblock {GAGE}: generally applicable gene set enrichment for pathway
  analysis.
\newblock {\em BMC Bioinformatics\/}, {\bf 10}, 161.

\bibitem[Mootha {\em et~al.}(2003)Mootha, Lindgren, Eriksson, Subramanian,
  Sihag, Lehar, Puigserver, Carlsson, Ridderstr{\aa}le, Laurila, Houstis, Daly,
  Patterson, Mesirov, Golub, Tamayo, Spiegelman, Lander, Hirschhorn, Altshuler,
  and Groop]{MLES03}
Mootha, V.~K. {\em et~al.} (2003).
\newblock {PGC-1alpha-responsive} genes involved in oxidative phosphorylation
  are coordinately downregulated in human diabetes.
\newblock {\em Nat Genet\/}, {\bf 34}(3), 267--273.

\bibitem[Press {\em et~al.}(2007)Press, Teukolsky, Vetterling, and
  Flannery]{PTVW07}
Press, W.~H. {\em et~al.} (2007).
\newblock {\em Numerical Recipes 3rd Edition: The Art of Scientific
  Computing\/}.
\newblock Cambridge University Press, 3 edition.

\bibitem[Sharma {\em et~al.}(2009)Sharma, Weber, Bairlein, Greff, K{\'e}ri,
  Cox, Olsen, and Daub]{SWBG09}
Sharma, K. {\em et~al.} (2009).
\newblock Proteomics strategy for quantitative protein interaction profiling in
  cell extracts.
\newblock {\em Nat Methods\/}, {\bf 6}(10), 741--744.

\bibitem[Smid and Dorssers(2004)Smid and Dorssers]{SD04}
Smid, M. and Dorssers, L. C.~J. (2004).
\newblock {GO-Mapper}: functional analysis of gene expression data using the
  expression level as a score to evaluate gene ontology terms.
\newblock {\em Bioinformatics\/}, {\bf 20}(16), 2618--2625.

\bibitem[Stojmirovi{\'c} and Yu(2007)Stojmirovi{\'c} and Yu]{SY07}
Stojmirovi{\'c}, A. and Yu, Y.-K. (2007).
\newblock Information flow in interaction networks.
\newblock {\em J Comput Biol\/}, {\bf 14}(8), 1115--1143.

\bibitem[Stojmirovi{\'c} and Yu(2009)Stojmirovi{\'c} and Yu]{SY09}
Stojmirovi{\'c}, A. and Yu, Y.-K. (2009).
\newblock {ITM Probe}: analyzing information flow in protein networks.
\newblock {\em Bioinformatics\/}, {\bf 25}(18), 2447--2449.

\bibitem[Subramanian {\em et~al.}(2005)Subramanian, Tamayo, Mootha, Mukherjee,
  Ebert, Gillette, Paulovich, Pomeroy, Golub, Lander, and Mesirov]{STMM05}
Subramanian, A. {\em et~al.} (2005).
\newblock Gene set enrichment analysis: a knowledge-based approach for
  interpreting genome-wide expression profiles.
\newblock {\em Proc Natl Acad Sci USA\/}, {\bf 102}(43), 15545--15550.

\bibitem[Sultan {\em et~al.}(2008)Sultan, Schulz, Richard, Magen, Klingenhoff,
  Scherf, Seifert, Borodina, Soldatov, Parkhomchuk, Schmidt, O'Keeffe, Haas,
  Vingron, Lehrach, and Yaspo]{SSRM08}
Sultan, M. {\em et~al.} (2008).
\newblock A global view of gene activity and alternative splicing by deep
  sequencing of the human transcriptome.
\newblock {\em Science\/}, {\bf 321}(5891), 956--960.

\bibitem[Wood {\em et~al.}(1993)Wood, Booth, and Butler]{WBB93}
Wood, A. T.~A. {\em et~al.} (1993).
\newblock Saddlepoint approximations to the cdf of some statistics with
  nonnormal limit distributions.
\newblock {\em Journal of the American Statistical Association\/}, {\bf
  88}(422), 680--686.

\end{thebibliography}

\newpage
\setcounter{figure}{0} 
\setcounter{table}{0} 

\appendix

\section{Saddlepoint approximation of tail probabilities}

References about saddlepoint approximations of the tail probabilities of random variables are abundant \cite{LR80,Daniels87,WBB93,Jensen95}. For completeness of our exposition we here present the derivation of the Lugannani-Rice formula \cite{LR80}, relying extensively on expositions by Daniels \cite{Daniels87} and Woods, Booth and Butler \cite{WBB93}.

Let $X$ be a continuous random variable supported on a subset of $\R$. We will assume that its probability density function (PDF), denoted by $f_X$ exists and that its moment generating function (MGF), defined by $\rho_X(t)=\int_{-\infty}^\infty f_X(x) e^{tx}\,dx$ converges for real $t\in [a,b]$ where $a<0<b$. Recall that $\rho_X(it)$ gives the characteristic function of $X$, that is, the Fourier transform of $f_X$ and that $f_X$ can hence be recovered by the Fourier inversion formula:
\begin{align}
f_X(x) & = \frac{1}{2\pi}\int_{-\infty}^\infty e^{-itx} \rho_X(it)\, dt\\
& = \frac{1}{2\pi}\int_{-\infty}^\infty e^{K_X(it)-itx}\, dt\\
& = \frac{1}{2\pi i}\int_{-i\infty}^{i\infty} e^{K_X(t)-tx}\, dt,
\end{align}
where $K_X(t)=\ln\rho_X(t)$ denotes the cumulant generating function (CGF) of $X$. The tail probability or P-value for a value $y$ (with respect to $X$), which we will denote by $Q_X(y)$ can be expressed as 
\begin{align}
Q_X(y) & = \prob(X\geq y) = \int_{y}^\infty f_X(x)\, dx\\
&= \frac{1}{2\pi i}\int_{y}^{\infty}\int_{-i\infty}^{i\infty} e^{K_X(t)-tx}\, dt\, dx\\
&= \frac{1}{2\pi i}\int_{-i\infty}^{i\infty}\int_{y}^{\infty} e^{K_X(t)-tx}\, dx\, dt\\
&= \frac{1}{2\pi i}\int_{c-i\infty}^{c+i\infty} e^{K_X(t)-ty}\, \frac{dt}{t}, \label{eq:inv_formula1}
\end{align}
where $c\in(0,b)$ is a constant introduced to avoid the pole at $t=0$.

Let $S$ denote the sum of $m$ independent, identically distributed random variables. We write $S=\sum_{j=1}^m X_j$, where $f_{X_j}=f_X$ for all $j$. Our goal is to derive an asymptotic approximation for the tail probability $Q_S$. It can be easily shown that $\rho_S(t)=\rho^m_X(t)$ and hence by \refeq{eq:inv_formula2} 
\begin{equation} \label{eq:inv_formula2}
Q_S(s) = \frac{1}{2\pi i}\int_{c-i\infty}^{c+i\infty} e^{mK_X(t)-ts}\, \frac{dt}{t}.
\end{equation}
To produce our approximation we note that the main contributions to the integral \refeq{eq:inv_formula2} occur in the neighborhood of the pole at $t=0$ and in the neighborhood of the saddle point $t=\hat{\lambda}$ where the exponent $I(t)=mK_X(t)-ts$ has a maximum, that is, where $I'(t)=0$. The saddlepoint condition is thus
\begin{equation}\label{eq:saddle1}
s = mK'_X(\hat{\lambda}),
\end{equation}
or alternatively 
\begin{equation}\label{eq:saddle2}
\int_{-\infty}^\infty \left(x-\frac{s}{m}\right) f_X(x) e^{\hat{\lambda}x}\,dx = 0.
\end{equation}
Let $\E(X)$ denote the expectation of $X$. Daniels \cite{Daniels54} has shown that eq.\,\refeq{eq:saddle1} has a unique simple root under most conditions. The value of $\hat\lambda$ increases with $s$, with $\sgn(\hat\lambda) = \sgn(s-m\E(X))$.

When $s\gg m\E(X)$, the contribution of the pole at $t=0$ to \refeq{eq:inv_formula2} is very small and to obtain an asymptotic approximation to $Q_S$ one can proceed by expanding $I(t)$ as a Taylor's series about $t=\hat\lambda$ and integrating the resulting integral term-by-term \cite{Daniels87}. However, as $s$ gets closer to the mean $\E(S)=m\E(X)$, such approximation performs poorly and in fact is unbounded at the mean. The essence of the method of \cite{Bleistein66} as applied to $Q_S$ by Lugannani and Rice \cite{LR80} is to produce a transformation of the integral \refeq{eq:inv_formula2} that would take into account the pole and hence to produce an approximation uniformly valid over the whole range of $S$. 

Make a transformation from $t$ to a new variable $z$ by
\begin{equation}\label{eq:transform1}
K_N(z)-\hat{z}z = mK_X(t) - ts,
\end{equation}
where $N$ denotes the Gaussian random variable with PDF $f_N(x)=\phi(x)=\exp(-x^2/2)/\sqrt{2\pi}$ and $Q_N(x)=\Phi(x)=\int_x^\infty\phi(t)\,dt$ and $s$ satisfies \refeq{eq:saddle1}. The value $\hat{z}$ is chosen so that the minimum of the left side is equal to the minimum of the right side, which occurs when $t=\hat{\lambda}$. Since $K_N(z)=\frac{1}{2}z^2$, eq. \refeq{eq:transform1} becomes
\begin{equation}\label{eq:transform2}
\frac{1}{2}z^2-\hat{z}z = mK_X(t) - tmK'_X(\hat\lambda).
\end{equation}
To find $\hat{z}$, we set $z=\hat{z}$ and $t=\hat\lambda$ in \refeq{eq:transform2} to get
\begin{equation}\label{eq:transform3}
-\frac{1}{2}\hat{z}^2 = m(K_X(\hat\lambda) - \hat\lambda K'_X(\hat\lambda))
\end{equation}
or, taking the sign for $\hat{z}$ to be equal to the sign of $\hat\lambda$,
\begin{align}
\hat{z} & = \sgn(\hat\lambda) \sqrt{(2m(\hat\lambda K'_X(\hat\lambda)-K_X(\hat\lambda))}\\
& = \sgn(\hat\lambda)\sqrt{(2(\hat\lambda s-mK_X(\hat\lambda))} \label{eq:hatz2}.
\end{align}

The transformation \refeq{eq:transform2} maps the region $[0,\hat\lambda]$ in $t$-space into the region $[0,\hat{z}]$ in $z$-space. The local behavior of $mK_X(t) - tmK'_X(\hat\lambda)$, which vanishes at $t=0$ and has zero derivative at $t=\hat\lambda$ is reproduced by $\frac{1}{2}z^2-\hat{z}z$ with similar behavior at $z=0$ and $z=\hat{z}$. Let $u=z-\hat{z}$. Then,
\begin{equation}
\frac12 u^2 = mK_X(t) - ts - mK_X(\hat\lambda) + \hat\lambda s.
\end{equation}
Expanding $mK_X(t) - ts$ about $t=\hat\lambda$ we have
\begin{align}
\frac12 u^2 &= \frac12 mK''_X(\hat\lambda)v^2 + \frac16 mK'''_X(\hat\lambda)v^3 + \ldots \\
& = \frac12 mK''_X(\hat\lambda)v^2 \left(1+\alpha_3 v + \alpha_4 v^2 + \ldots \right)
\end{align}
where $v=t-\hat\lambda$ and $\alpha_n=\frac{2K^{(n)}_X(\hat\lambda)}{n! K''_X(\hat\lambda)}$. Hence,
\begin{equation}\label{eq:expansion1}
u = \sqrt{mK''_X(\hat\lambda)} v\left(1+\alpha_3 v + \alpha_4 v^2 + \ldots \right)^{1/2}.
\end{equation}
It follows that $du/dv$ and $dv/du$ are nonzero for all $v\in[0,\hat\lambda]$ and $u\in[0,\hat{z}]$, respectively. Since $K_X$ is analytic in the region of interest, $u(v)$ and $v(u)$ are analytic over the same intervals. Obviously, the same conclusion follows for $z$ as a function of $t$ and $t$ as a function of $z$. By the inverse function theorem, the transformation $t\leftrightarrow z$ can be extended to a bijection between complex neighborhoods of $[0,\hat\lambda]$ and $[0,\hat{z}]$.

The integral \refeq{eq:inv_formula2} now transforms (using Cauchy's theorem) into  
\begin{equation} \label{eq:inv_formula3}
Q_S(s) = \frac{1}{2\pi i}\int_{d-i\infty}^{d+i\infty} e^{K_N(z)-z\hat{z}}\, \left(\frac{1}{t}\frac{dt}{dz}\right)\, dz,
\end{equation}  
where $d>0$. For small $t$, we can write 
\begin{equation}
z \approx z|_{t=0} + t\frac{dz}{dt}\Big\vert_{t=0} = t\frac{dz}{dt}\Big\vert_{t=0}.
\end{equation}
When $\hat\lambda\neq 0$ and hence $\hat{z}\neq 0$, differentiating \refeq{eq:transform2} we obtain
\begin{equation}\label{eq:dtransform2}
\frac{dz}{dt} = \frac{mK'_X(t) - mK'_X(\hat\lambda)}{z-\hat{z}},
\end{equation}
while when $\hat\lambda= 0$, \refeq{eq:expansion1} implies $dz/dt=du/dv \approx \sqrt{mK''_X(0)}$ when $t$ is small. Thus,
\begin{equation}
\frac{dz}{dt}\Big\vert_{t=0} = 
\begin{cases}
\frac{1}{\hat{z}}(s-m\E(X)) & \text{if $\hat\lambda\neq 0$},\\
\sqrt{mK''_X(0)} & \text{if $\hat\lambda=0$}
\end{cases}
\end{equation}
and therefore, for small $t$, $z\approx Ct$ where $C$ is a constant. Let 
\begin{equation}
U(z) = \left(\frac{1}{t}\frac{dt}{dz} - \frac{1}{z}\right).
\end{equation}
By expanding $mK_X(t) - ts$ about $t=0$, it can be shown that, $\lim_{z\to 0} U(z) < \infty$ and, since $dt/dz$ is analytic, $U(z)$ is analytic in the neighborhood of $z=0$ that includes $\hat{z}$. Therefore, we can rewrite the integral \refeq{eq:inv_formula3} as
\begin{align} 
Q_S(s) &= \frac{1}{2\pi i}\int_{d-i\infty}^{d+i\infty} e^{K_N(z)-z\hat{z}}\, \frac{dz}{z} \label{eq:inv_formula4} \\
& + \frac{1}{2\pi i}\int_{d-i\infty}^{d+i\infty} e^{K_N(z)-z\hat{z}}\, U(z)\, dz \label{eq:inv_formula5}.
\end{align}
The singularity has now been isolated into \refeq{eq:inv_formula4}, which, by comparing with \refeq{eq:inv_formula2}, we recognize to equal $\Phi(\hat{z})$. On the other hand,
$U(z)$ can be expanded as a Taylor's series around the saddlepoint $z=\hat{z}$ and integrated to obtain an asymptotic series for \refeq{eq:inv_formula5}. For the first-order approximation, that is, the leading behavior, we only take the constant term at $\hat{z}$. Let
\begin{align}
\hat{y} = t\frac{dz}{dt}\Big |_{t=\hat\lambda} =  \hat\lambda\frac{du}{dv}\Big |_{v=0} =\hat\lambda\sqrt{mK''_X(\hat\lambda)}.
\end{align}
Then, $U(\hat{z})=1/\hat{y} - 1/\hat{z}$ and the integral \refeq{eq:inv_formula5} becomes
\begin{equation}
U(\hat{z})\frac{1}{2\pi i}\int_{\hat{z}-i\infty}^{\hat{z}+i\infty} e^{K_N(z)-z\hat{z}}\, dz = \left(\frac{1}{\hat{y}}- \frac{1}{\hat{z}} \right)\phi(\hat{z}).
\end{equation}
Thus, we have obtained the Lugananni-Rice formula:
\begin{equation}\label{eq:tail}
\prob(S\geq s) = \Phi(\hat{z}) + \left(\frac{1}{\hat{y}}- \frac{1}{\hat{z}} \right) \phi(\hat{z}),
\end{equation}
with $\hat{z}(s)$ given by \refeq{eq:saddle1} and \refeq{eq:hatz2}.

\section{SaddleSum implementation}

As mentioned in the main text, our SaddleSum algorithm approximates term P-values by first solving eq.\,\refeq{eq:saddle1} for $\hat{\lambda}$ using Newton's method and then using the Lugananni-Rice formula \refeq{eq:tail}. The key step is estimation of $\hat\lambda$. Since the moment-generating function $\rho$  of the underlying space $W$ is not known, we estimate it (and its derivatives) using $\vec{w}$. Given sufficiently many weights ($n >> 1$), the results can be quite accurate (see below). One limitation of this approach is that our approximation can only accept scores not greater than $m$ times maximal weight ($\hat\lambda$ becomes infinite at this bound). Thus, the approximation can be inaccurate for very large scores, causing a larger than usual relative error in P-values (Fig.~S6). However, occurence of such extreme scores is rarely seen in practice.

Theoretically, Lugananni-Rice formula is valid over the whole range of the distribution, for small and large scores and both near the mean and in the tails \cite{LR80}. However, the form \refeq{eq:tail} becomes numerically unstable close to the mean of the distribution (i.e. when $\hat\lambda$ is close to 0). Alternative asymptotic approximations exist that are numerically stable near the mean \cite{Daniels87}. For SaddleSum, we were mainly interested in the tail probabilities and we therefore decided not to attempt to approximate the P-values of the scores smaller than one standard deviation from the mean (SaddleSum returns P-value of 1 for all such scores). Terms with such scores are never significant in the context of enrichment analysis.

When processing a terms database, we retain previously computed values of $\hat\lambda$ with associated scores and parameters for Lugananni-Rice formula in a sorted array. Since $\hat\lambda$ and the P-value are monotonic with respect to the score, using binary search we can certify for many terms that their P-value is larger than a given cutoff and hence eliminate them without running Newton's method. Furthermore, binary search provides a bracket for $\hat\lambda$ and hence Newton's method usually converges in very few iteration. We use the bracketed version of Newton's method recommended in the Numerical Recipes book \cite{PTVW07} (Section 9.4). This combines the classical Newton's method with bisection and has guaranteed global convergence.

We show evaluations of SaddleSum performance against some theoretically well-characterized distributions in Fig.~S5 and S6. It can be seen that the relative error between the SaddleSum approximation and the theoretical P-value is generally very small except for extremely large scores, when P-values are very small. In the context of the enrichment analysis, this discrepancy is not important because such terms will be evaluated as highly significant even if the P-value is off by few orders of magnitude. To further illustrate the quality of our approximation, we have computed the Kullback-Leibler (KL) divergences (relative entropies) between the tail distribution implied by SaddleSum and the theoretical distribution. Prior to computation of KL divergence, both distributions were normalized over the region where SaddleSum is valid (i.e. the tail with scores larger than one standard deviation over the mean). All KL divergence values are extremely small and are comparable between distributions.

Fig.~S7 shows relative errors of SaddleSum compared to the empirical distributions using the same weights and term sizes as for Fig.~S1 and S2. In this case however, in agreement with the null model of SaddleSum, we sampled weights with replacement. Our results indicate that, except for small $m$ with weights coming from network flow simulations, the relative error of the SaddleSum is similar to that obtained in comparison with well-characterized distributions.

\makeatletter
\renewcommand\section{\@startsection{section}{1}{\z@}%
                                    {0\p@}{0\p@}
                                    {\Large\bfseries\phantom}}
\makeatother
\section{Supplementary figures}

\newpage
\begin{figure*}
\begin{center}
\includegraphics{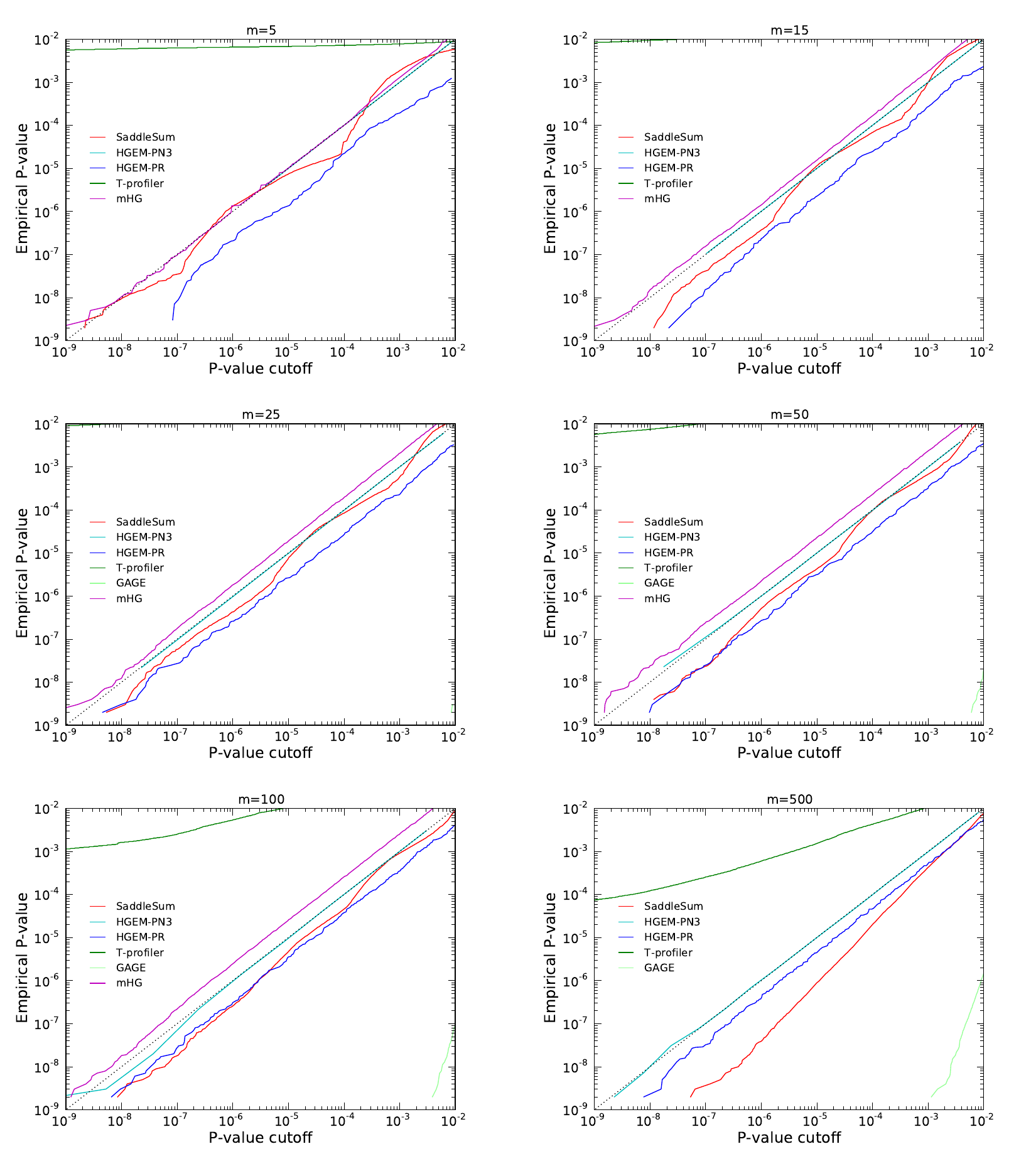}
\end{center}
\renewcommand{\thefigure}{S\arabic{figure}}
\caption{{\small Accuracy of reported P-values from simulations using weights from 100 results of protein network information flow simulations. Each graph shows empirical P-values associated with reported P-value cutoffs for investigated enrichment methods, obtained from queries of decoy term datasets with fixed size terms. The curves for GAGE are omitted from the plots for term sizes 5, 15 and 25 because all reported P-values were greater than $10^{-2}$. The graph for $m=500$ misses the results for mHG because we could not finish the simulation runs within any reasonable amount of time.}}
\end{figure*}

\newpage
\begin{figure*}
\begin{center}
\includegraphics{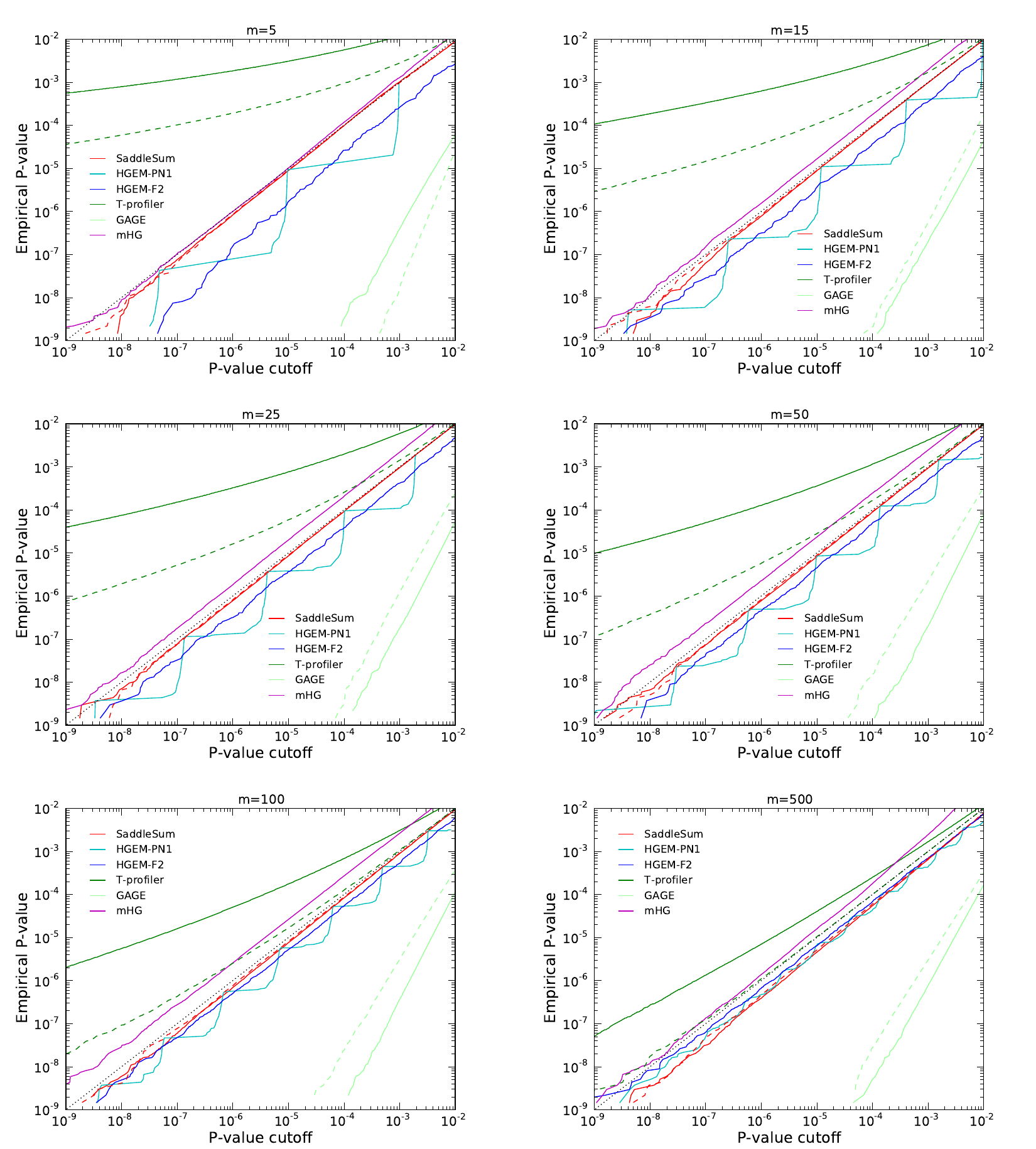}
\end{center}
\renewcommand{\thefigure}{S\arabic{figure}}
\caption{{\small Accuracy of reported P-values from simulations using weights from 136 microarrays. Each graph shows empirical P-values associated with reported P-value cutoffs for investigated enrichment methods, obtained from queries of decoy term datasets with fixed size terms. For SaddleSum, T-profiler and GAGE, full lines indicate the results where negative weights were set to 0, while dashed lines show the results using all weights.}}
\end{figure*}

\newpage
\begin{figure*}
\begin{center}
\includegraphics[scale=0.85]{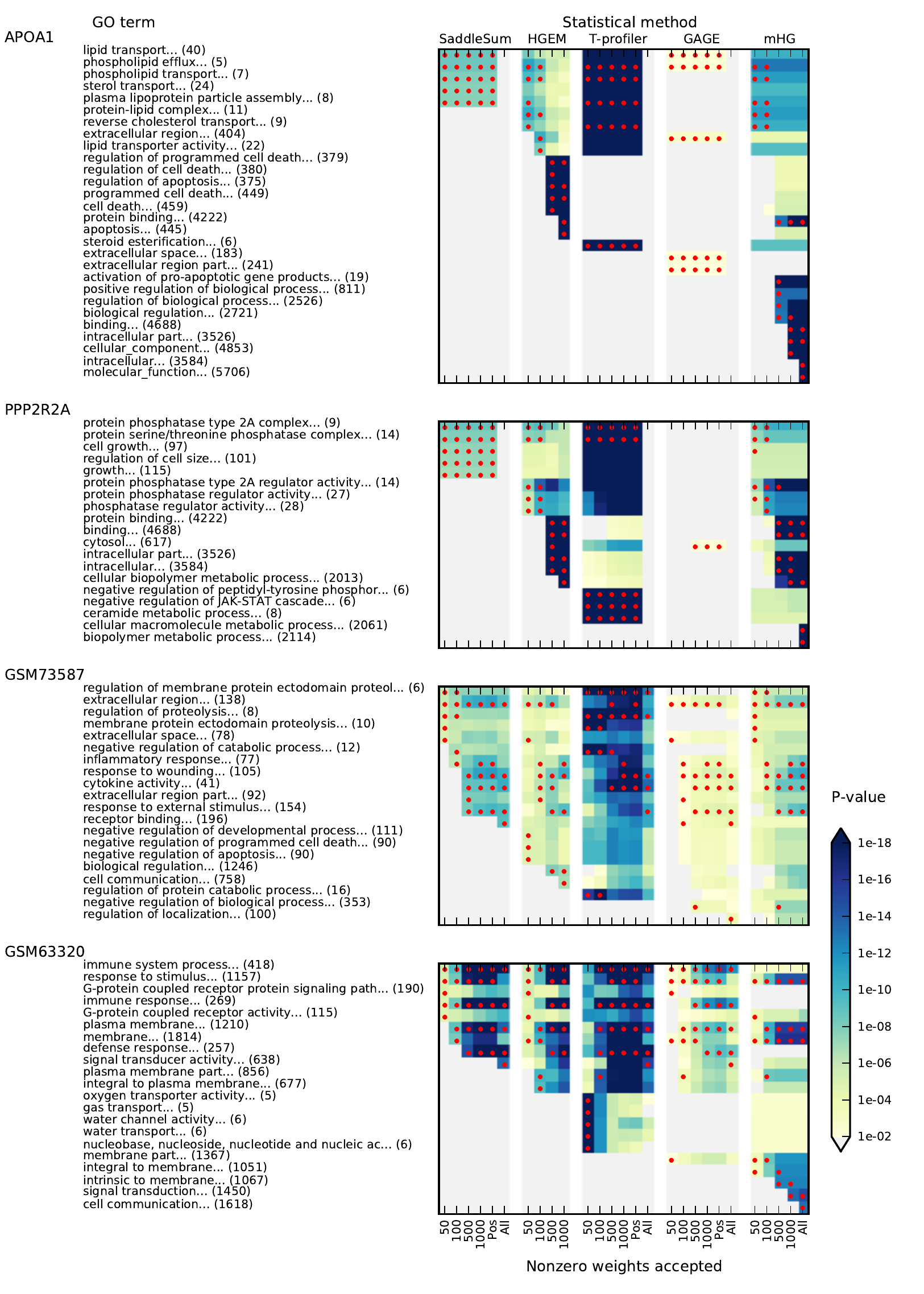}
\end{center}
\renewcommand{\thefigure}{S\arabic{figure}}
\caption{{\small Additional examples of sets of top-five GO terms retrieved by evaluated methods (refer to Fig.\,2B for full explanation.) The upper two panels show the enrichment results using the weights from outputs of \ITM\ emitting mode with human proteins APOA1 (apolipoprotein A-I, a major protein component of high density lipoprotein in plasma) and PPP2R2A (phosphatase 2 regulatory subunit B) as sources. The lower two panels show the results using weights from microarrays investigating mast cell activation (GSM73587) and malaria response (GSM63320).}}
\end{figure*}

\newpage
\begin{figure*}
\begin{center}
\includegraphics{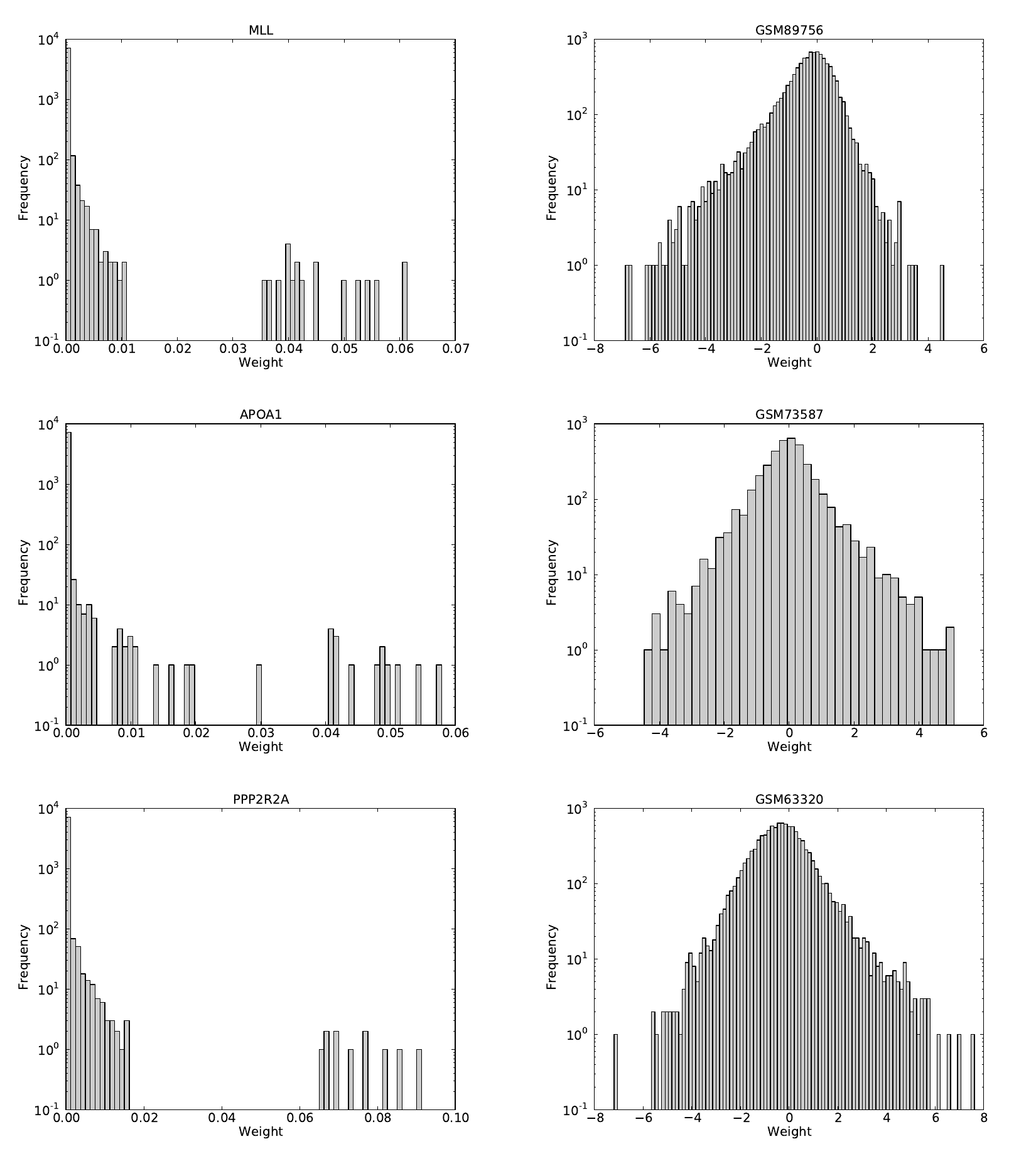}
\end{center}
\renewcommand{\thefigure}{S\arabic{figure}}
\caption{{\small Distributions of weights for examples from Fig.\,2 and Fig.\,S3. Network examples are shown on the left, microarray on the right.}}
\end{figure*}

\newpage
\begin{figure*}
\begin{center}
\includegraphics{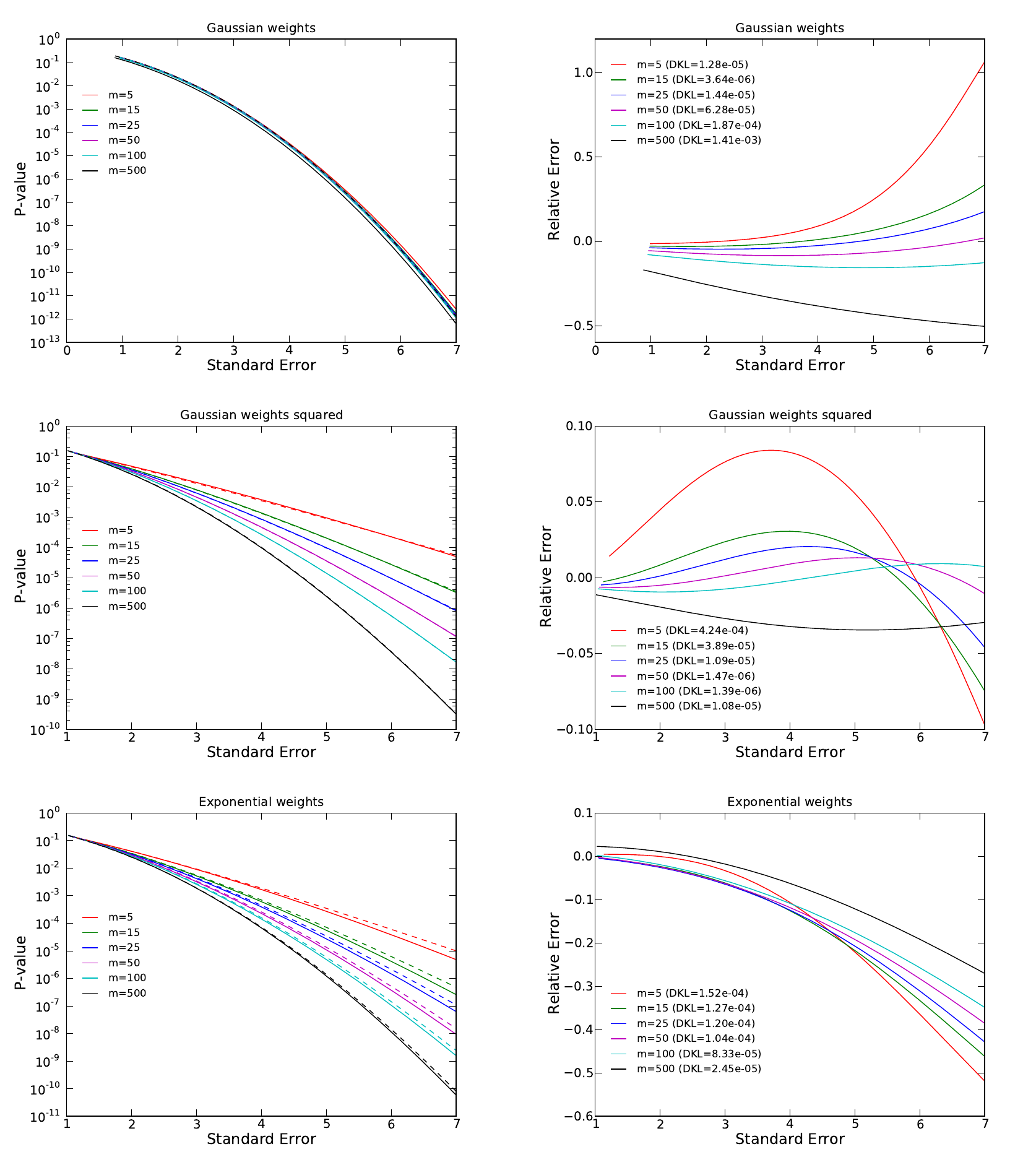}
\end{center}
\renewcommand{\thefigure}{S\arabic{figure}}
\caption{{\small P-values (left) and relative errors (right) for SaddleSum approximations of sums of i.i.d. continuous random variables that are characterized theoretically. In each case 10000 weights were randomly sampled from a distribution and used as input to SaddleSum. The P-values from SaddleSum were compared with P-values from theoretical distributions of the sum of $m$ numbers. Kullback-Leibler divergences (DKL) between the approximated tails of distributions are shown in parentheses for each $m$. Top: Gaussian (standard normal) weights -- sum follows normal distribution. Middle: squared Gaussian weights -- sum follows Chi-squared distribution. Bottom: weights from exponential distribution -- sum follows Erlang distribution. }} 
\end{figure*}

\newpage
\begin{figure*}
\begin{center}
\includegraphics{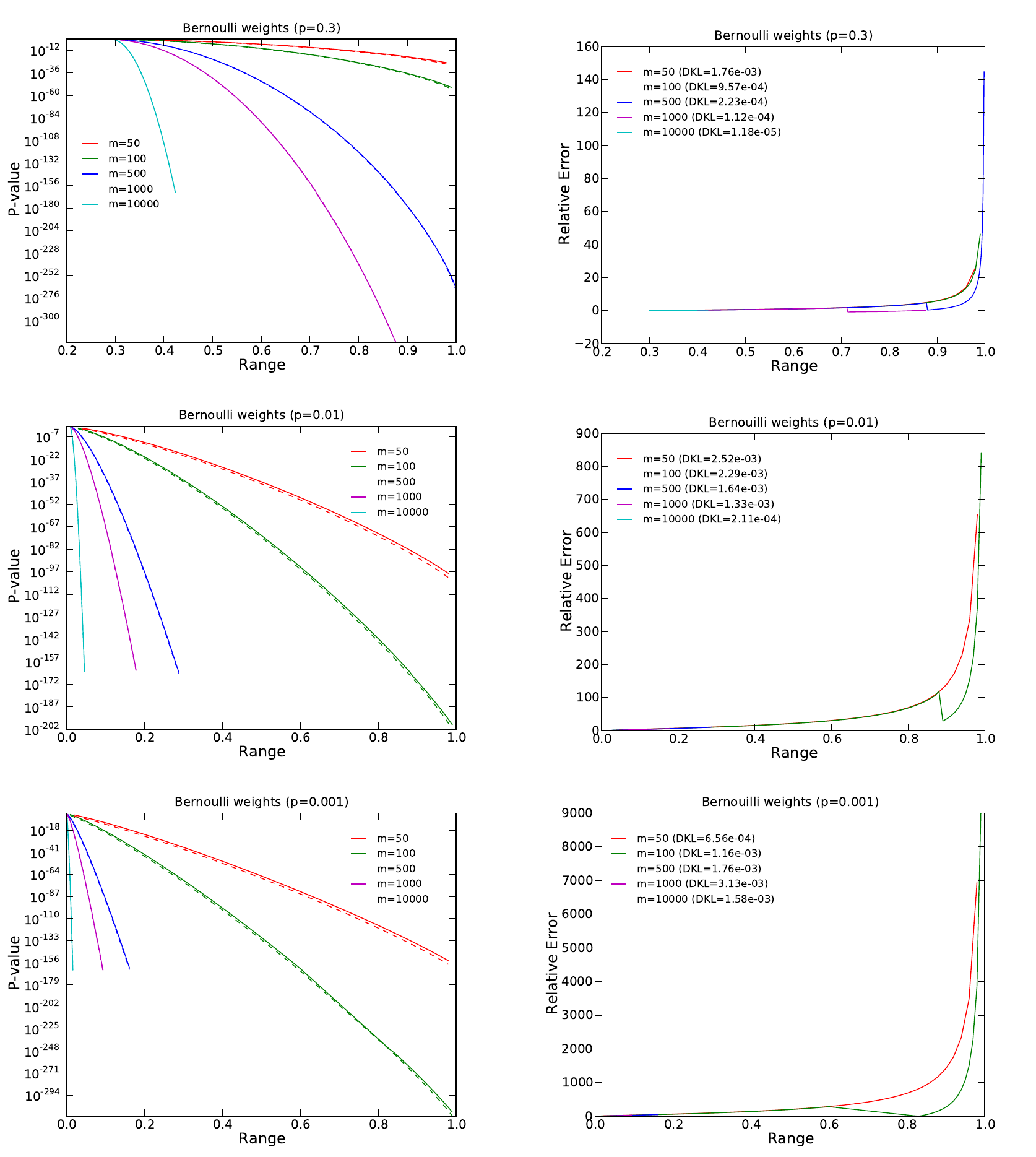}
\end{center}
\renewcommand{\thefigure}{S\arabic{figure}}
\caption{{\small P-values (left) and relative errors (right) for SaddleSum approximations of sums of i.i.d. Bernoulli ($\{0,1\}$) random variables with different parameter $p$. Such sums follow binomial distribution. In each case 10000 weights were randomly sampled from a distribution and used as input to SaddleSum. The P-values from SaddleSum were compared with P-values from the binomial distribution. Kullback-Leibler divergences between the approximated tails of distributions are shown in parenthesis for each $m$. Top: $p=0.3$. Middle: $p=0.01$. Bottom: $p=0.001$. The dramatic increase in relative error is caused by $\hat\lambda$ instability at extreme scores, see Section B.}}
\end{figure*}

\newpage
\begin{figure*}
\begin{center}
\includegraphics{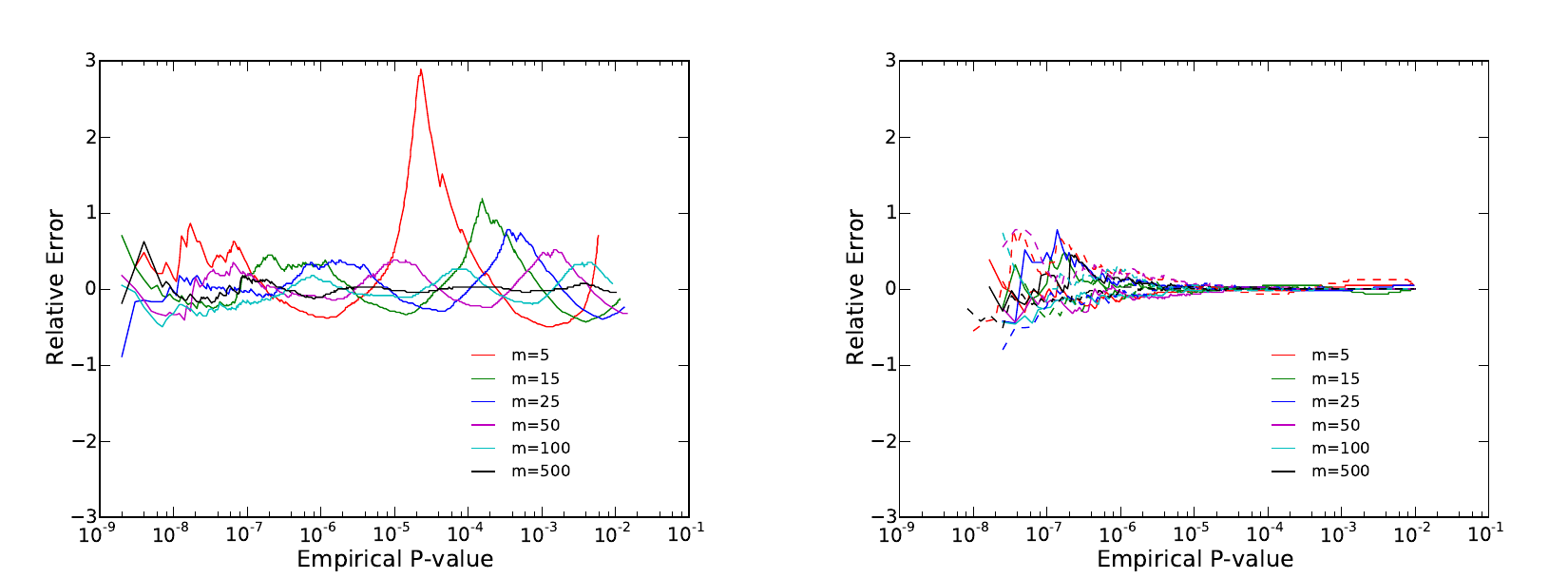}
\end{center}
\renewcommand{\thefigure}{S\arabic{figure}}
\caption{{\small Relative error of P-values reported by SaddleSum from simulations using weights from 100 results of protein network information flow simulations (left) and from 136 microarrays (right). These are the same query sets as evaluated in Fig.~S1 and Fig.~S2 but in this case the weights are drawn with replacement. Each sample size $m$ is shown in different color. Full lines indicate the results where negative weights were set to 0, while dashed lines show the results using all weights.}}
\end{figure*}
\end{document}